\documentclass[%
 reprint,
 amsmath,amssymb,
 aps,
]{revtex4-1}

\usepackage{graphicx}% Include figure files
\usepackage{dcolumn}% Align table columns on decimal point
\usepackage{bm}% bold math
\begin{document}

%\preprint{APS/123-QED}

\title{Social Network Differences of Chronotypes Identified from Mobile Phone Data}% Force line breaks with \\

\author{Talayeh Aledavood}

\affiliation{Department of Computer Science, Aalto University, FI-00076 AALTO, Espoo, Finland}%


\author{Sune Lehmann}
\affiliation{
DTU Compute, Technical University of Denmark, Kgs. Lyngby, Denmark}

\affiliation{
The Niels Bohr Institute, University of Copenhagen, Copenhagen, Denmark}
\author{Jari Saram\"aki}
\affiliation{%
Department of Computer Science, Aalto University, FI-00076 AALTO, Espoo, Finland
}%


\date{\today}% It is always \today, today,

\begin{abstract}
Human activity follows an approximately 24-hour day-night cycle, but there is significant individual variation in awake and sleep times. Individuals with circadian rhythms at the extremes can be categorized into two chronotypes: ``larks'', those who wake up and go to sleep early, and ``owls'', those who stay up and wake up late. It is well established that a person's chronotype can affect their activities and health. However, less is known on the effects of chronotypes on the social behavior, even though it is evident that social interactions require coordinated timings. To study how chronotypes relate to social behavior, we use data collected using a smartphone app on a population of more than seven hundred volunteer students to simultaneously determine their chronotypes and social network structure. We find that owls maintain larger personal networks, albeit with less time spent per contact. On average, owls are more central in the social network of students than larks, frequently occupying the dense core of the network. Owls also display strong homophily, as seen in an unexpectedly large number of social ties connecting owls to owls.

\end{abstract}

\pacs{Valid PACS appear here}% PACS, the Physics and Astronomy
\maketitle

%\tableofcontents

\section{\label{sec:intro}Introduction}

Life on Earth follows a circadian rhythm~\cite{panda2002circadian, dibner2010mammalian, czeisler1999stability, edery2000circadian}. This circadian pattern includes human activities and sleep, with rhythms reflected at the psychological, physiological, and biochemical levels~\cite{czeisler1980human, edgar2012peroxiredoxins}. Even though all humans are diurnal and endogenously controlled by an internal circadian clock, there are individual differences in how the internal clock is synced with the environment's daily rhythm~\cite{horne1977InterIndividual, kerkhof1985InterIndividual}. 
These differences can be classified with three \emph{chronotypes}~\cite{adan2012circadian}.
At the two extremes are the morning-active people (``larks'') and the evening-active people (``owls''), and the rest fall in the intermediate category whose rhythms do not deviate much from the population average (note that there are no absolute criteria for any given chronotype). It has been shown that a person's chronotype can change over the course of her life, but it is fairly stable within time periods of the order of a few years~\cite{lee2011ChronoChange}. 
 
In the past two decades there has been a lot of interest in the epidemiology of chronotypes~\cite{roenneberg2007epidemiology} in terms of \emph{e.g.}~age, gender, personality, income, or health risks~\cite{tankova1994circadian, adan2012circadian, roenneberg2016circadian}. Chronotypes have been argued to correlate with certain personal or behavioral traits; these include sociosexual orientation \cite{jankowski2014}, personality (see, \emph{e.g.}, ~\cite{adan2012circadian}), academic performance~\cite{preckel2011chronotype}, body mass index~\cite{yu2015evening, arora2015associations}, physical and mental health~\cite{gaspar2009, wong2015, roenneberg2016circadian, merikanto2014evening}, and how people make use of their time~\cite{kauderer2013}. 
However, even though it is known that circadian rhythms and chronotypes have an important social component, there has been less emphasis on the \emph{sociology} of chronotypes, that is, how different chronotypes relate to the structure of social systems. It is known that social cues are important \emph{zeitgebers} (``time-givers''), affecting the phases of circadian rhythms~\cite{roenneberg2007epidemiology}; further, social interactions require some level of synchronization of activities. Therefore, it is reasonable to expect that chronotypes and social network structure are somehow correlated.

\begin{figure*}[!tb]
\centering
\includegraphics[width=.75\linewidth]{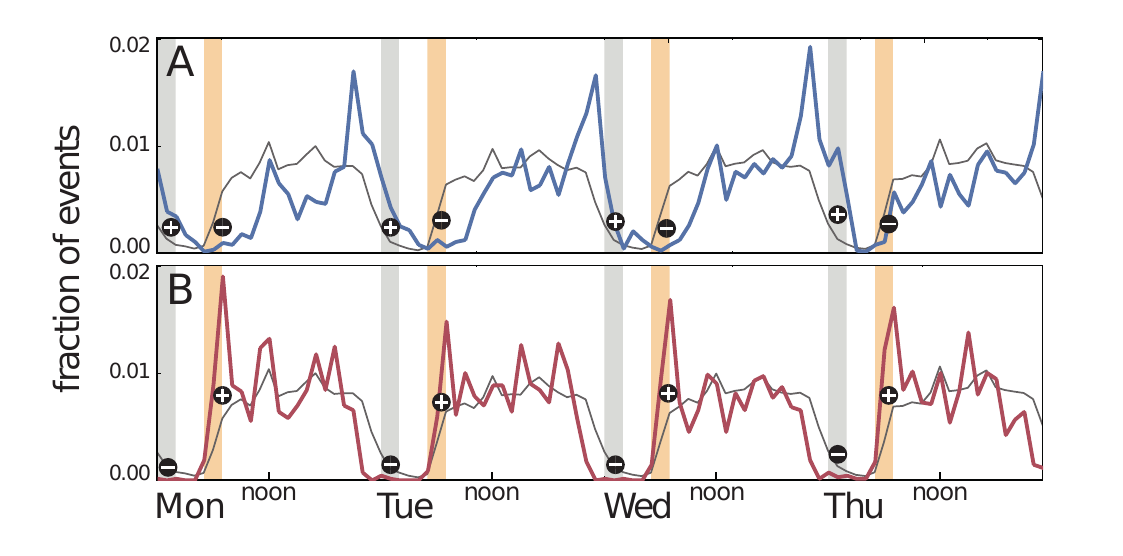}
\caption{The chronotypes of participants are determined by computing the average hourly frequencies of the `screen-on' events of their mobile phones for the first four weekdays. Each participant's pattern is compared to the population average (solid gray line). For chronotype identification, the gray and yellow time ranges are used. If a participant has above-average levels of activity (plus sign) in the night-time gray time range and below-average levels of activity (minus sign) in the yellow morning time range, she is categorized as an evening-active owl (panel A). Participants of the opposite pattern (below-average night-time activity and above-average morning activity) are categorized as morning-active larks. For details, see Materials and Methods.}
\label{fig:fig1_method}
\end{figure*}

% SUNE COMMENT
% One might argue (based on example A) that it would make sense to use wider periods than the gray/yellow time ranges.

In this paper, we set out to explore the relationship between chronotypes and social networks. In particular, we have two hypotheses i) social ties  display chronotype homophily, where ties are more frequent between individuals of the same chronotype, and that ii) different chronotypes inhabit different positions in social networks. The first hypothesis follows from the need to synchronize social interactions, and the second one results from the first when viewed from the whole network's point of view. To make progress, we take advantage of a subset (about $700$ individuals) of a rich dataset of digital activity amongst nearly $1000$ students, collected for over more than a year at the Technical University of Denmark (DTU). This set of data allows us to simultaneously assess the chronotypes of each student and reconstruct the structure of their social network. In this experiment, participants were equipped with identical smartphones and they volunteered to install an app that collected, among other things, detailed communication metadata as well as times of the phone screen turning on and off. See Materials and Methods for details and Ref.~\cite{stopczynski2014DTU} for a full description of the experiment and all recorded data.

Mobile phones have become an important part of most people's daily lives, and they are commonly used throughout the day. We take advantage of this fact in order to identify chronotypes. Rather than using one of the common questionnaires for identifying chronotypes~\cite{horne1975MEQ, roenneberg2003life} that may suffer from recall bias, we identify chronotypes based on phone usage. Our chronotype identification is based on data on the times when the participants' smartphone screens have been switched on (see Materials and Methods for details), using these events as a proxy of a person's activity level. Even though these data do not fully account for a person's activity, the frequency of screen-on events is seen to follow a typical $24\mathrm{h}$ cycle where longer periods of inactivity coincide with nights. Because there are clear and persistent individual differences in the levels of early-morning or late-night activity, we use them as a way to assess the chronotypes of individuals~\cite{aledavood2015daily, aledavood2015cycles}.

In order to reconstruct social networks, we use data on calls and text messages between the study participants, constructing a network where two participants $i$ and $j$ are linked if there is at least one call or text message from $i$ to $j$ and vice versa. For calculating the total personal network size of each participant, we also use calls and texts to (anonymized) persons outside the study cohort.

We find that chronotypes correlate with indicators of social behavior. Evening-active owls have larger personal networks than morning-active larks, albeit with less frequent contacts to each network member. From the perspective of the participants' entire social network, this translates to owls being more central than larks. They also display clear homophily with higher-than-randomly-expected numbers of ties connecting owls to owls. Surprisingly, this homophily is not visible in the case of larks.

\section*{Results}
\subsection{Screen-on events can be used for chronotype assessment}
We use time-stamped data on `screen-on' events from the smartphone data-collection apps to assign a behavioral chronotype to each participant. Whenever the participant uses the smartphone, from making a call to checking the time, the phone's screen is turned on, and the data-collection app records the time of this event. We use the frequency of these events as a statistical proxy for the daily activity rhythm of the participant, since frequent screen-on events provide information that the participant is awake, and night-time event frequencies are typically low or zero. To form an overview of the daily activity patterns of participants, we aggregate the screen-on event frequencies in hourly time bins for the four weekdays from Monday to Thursday for each of the $N=222$ participants who used their phones actively during our observation period (see Methods for details). A population-level average rhythm is computed for reference. 

\begin{figure}%[tbhp]
\centering
\includegraphics[width=.999\linewidth]{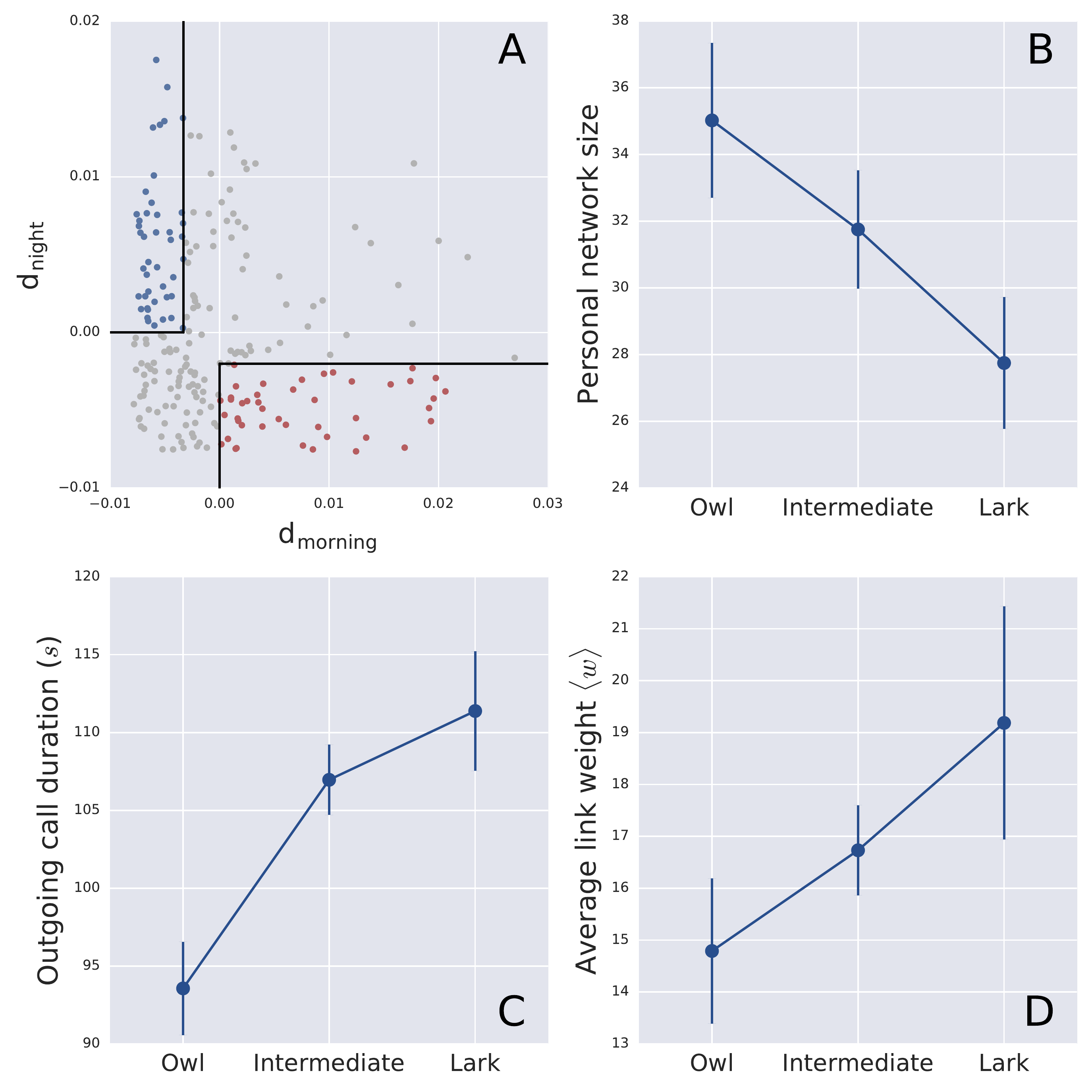}
\caption{Chronotypes and their behavioral differences. Panel A) displays the criteria for labeling participants as night-active owls (blue) and morning-active larks (red). $x$- and $y$-axes represent the difference between morning patterns and evening patterns of each individual's activity compared to the population average activity rhythm, respectively. Panels B, C, D: Average personal network size, average duration of outgoing calls, and average tie strength as measured by communication frequency, for the three different chronotypes.}
\label{fig:chronotype_behaviours}
\end{figure}

Figure~\ref{fig:fig1_method} displays the screen-on daily rhythms of two study participants (upper and lower panels), together with the population average. The phase of the daily pattern of the participant in panel A is consistently shifted towards the night, while the participant of panel B displays a pattern whose phase is shifted towards morning. These phase shifts are captured by the event frequencies in the early morning hours ($5$ AM to $7$ AM) and late hours of the day (midnight to $2$ AM); `larks' are associated with above-average morning activity and below-average night-time activity, and the opposite holds for `owls' (see Methods for details). On this basis, 20\% of the participants ($N=44$) are labeled as larks, $20\%$ as owls ($N=44$), and the rest as intermediate ($N=134$); the exact criteria have been set for obtaining these percentages (Methods). See panel A in Figure~\ref{fig:chronotype_behaviours}.

\subsection{Owls have larger personal networks than larks}
We first construct the personal networks of all participant based on both call and text data, using each (hashed and anonymized) phone number the participants call or text or get calls or texts from as a proxy of a social relationship. In this network, each individual is a node and communication events (calls and text messages) between people are the links. The degree of a node (personal network size) is the total number of people in contact with that node, while the strength is the number of times that a link is activated (total number of interactions between two nodes). When constructing the personal network of each individual, we consider outgoing and incoming calls and text messages to and from any phone number, not only those belonging to other study participants (see Methods). In addition to personal network membership, we count the total number of calls and texts with each contact, as well as the total call duration. We then study the properties of the personal networks of individuals of each chronotype separately. 

Figure~\ref{fig:chronotype_behaviours}B displays the average personal network sizes for students of each chronotype. It is evident that owls have personal networks that are much larger than those of larks, with the intermediate chronotype positioned in between (owls: network size $k=35.0 \pm 2.3$, larks: $k=31.7 \pm 1.8$, intermediate $k=27.7 \pm 2.0$). When the average call durations and total frequencies of calls or texts per social contact are considered, an opposite trend becomes visible (Fig.~\ref{fig:chronotype_behaviours} C and D): owls make the shortest calls on average and their communication frequency per social tie is the lowest as compared to the intermediate chronotype and in particular to larks. This reflects a known sub-linear scaling between node degree and strength in social networks (see, e.g.,~\cite{onnela2007b}); the larger the number of relationships, the less time is available for each of them. 

\subsection{Owls are more central than larks in the social network of participants}
In order to study the network centrality of each participant, we constructed the social network of participating students, so that two individuals $i$ and $j$ are connected with an unweighted link if there are either calls or text messages from $i$ to $j$ and from $j$ to $i$ (see Methods for details). This network consists of $N=734$ participants; out of these, $222$ had enough screen-on events to be assigned a chronotype (for filtering criteria, see Methods). 
We then computed the values of various network centrality measures for all individuals within each of the three chronotypes. The chosen measures were i) betweenness centrality, measuring the number of shortest paths through a network node, ii) closeness centrality, quantifying the inverted average geodesic distance to other nodes, iii) eigenvector centrality, reflecting the level of connectivity to high-centrality nodes in an iterative fashion, and iv) core number, indicating membership in a core where all nodes are linked to other member nodes with at least $k$ links. 

\begin{figure*}%[tbhp]
\centering
\includegraphics[width=.999\linewidth]{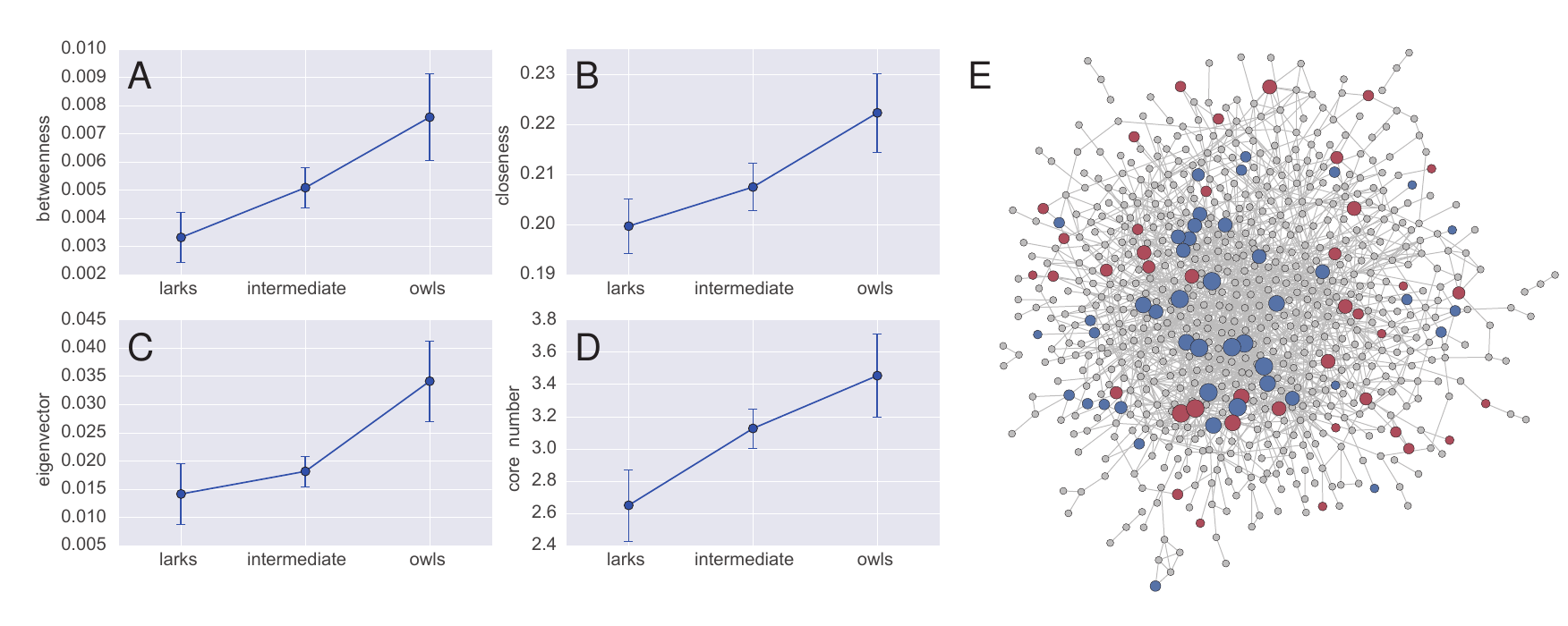}
\caption{Chronotypes are associated with network centrality such that individuals of the N chronotype are on average more central. Panels A-D: network centrality measures (betweenness centrality, closeness centrality, eigenvector centrality, and core number) for the three chronotypes. In all cases, values for the owl chronotype are the highest. Panel E: a network visualization of the social network of participants. Individuals of the owl (lark) chronotype are displayed as blue (red). The rest are gray. For blue and red nodes, the size is determined by core number. Nodes of the owl chronotype are more frequent in the central core of the network.}
\label{fig:network}
\end{figure*}

These four centrality measures are displayed in Fig.~\ref{fig:network} (panels A-D), together with a visualization of the network (panel E). There is an increasing trend in centrality from larks to owls for all centrality measures; the network visualization also clearly shows that owls (blue) are more frequently located in central parts of the network than larks (red). 

\subsection{Owls display strong homophily, whereas larks do not}
On the basis of the student network, we computed the number of links $E_{c}$ between individuals of each chronotype $c$ and the corresponding subnetwork edge density $\rho_c=2E_{c}/\left[N_c\left(N_c-1\right)\right]$, where $N_c$ is the number of individuals of each chronotype. These were compared to reference values computed for the null hypothesis where the existence of links is independent of the chronotypes of the connected persons: $E_{c,\mathrm{ref}}=EN_c\left(N_c-1\right)/ \left[N\left(N-1\right)\right]$, where $E$ and $N$ are the number of links and nodes in the whole network, respectively, and $\rho_{c,\mathrm{ref}}=2E/\left[N\left(N-1\right)\right]$. 

As seen in Table.~\ref{table:homophily}, the numbers of links and the link density between owls is far higher than randomly expected. This indicates strong homophily between owls. For larks, no such effect is observed. 

\begin{table*}[!ht]
\centering
\caption{Observed and expected numbers of links between members of the same chronotype and the respective network densities. The expectations are calculated using the null hypothesis that chronotypes are randomly located in the network. The $Z$-score measures how many standard deviations away the observed edge numbers are from the mean of the hypergeometric distribution corresponding to the random null hypothesis.}\label{sampletable}
\begin{tabular}{l l l l l l}
\hline
\textbf{Chronotype} & \textbf{\# of links $E$} & \textbf{Expected \# of links} $E_{H0}$ & \textbf{Subnetwork density} $\rho$ & \textbf{Expected density} $\rho_{H0}$ & \textbf{Z-score}\\
\hline
owls & 52 & 6.5 & 0.055 & 0.007 & 17.9 \\
intermediate & 120 & 61.4 & 0.013 & 0.007 & 7.6\\
larks & 5 & 6.5 & 0.007 & 0.005 & -0.6\\
\hline
\end{tabular}
\label{table:homophily}
\end{table*}

\section{Discussion}
There are well-known differences between human chronotypes that have been related to various personal and behavioral traits. In this study, we have focused on the social dimension of chronotypes, investigating how the chronotypes are reflected in the social network structure and position in social networks. This study is made possible by detailed behavioral data collected from volunteer students with smartphone data-collection apps in an experiment at the Technical University of Denmark. We defined chronotypes using `screen-on' event times and constructed the social network of students using mutual calls and text messages as proxies of social relationships. Our results establish three findings: there are significant differences in i) the personal network sizes of different chronotypes, ii) the network position of different chronotypes in terms of several network centrality measures that all yield similar results, and iii) the level of homophily between members of the same chronotype, with an asymmetry where evening-active owls display strong network homophily whereas larks do not.

This is the first study to investigate the relationship between chronotypes of individuals and the structure of their social networks, even though the role of social \emph{zeitgebers}~\cite{allebrandt2014} in synchronizing circadian rhythms is known and it is evident that social activities require a common understanding of their times. The closest results are related to correlations between the evening-active chronotype and extroversion (see ~\cite{adan2012circadian} for a review). One might reasonably expect that these correlations translate into properties of personal networks such that owls maintain larger communication networks. However, results from past research do not all agree that this is the case. 

Regarding our findings, it is well-known that the values of many network centrality measures correlate with node degrees and therefore the differences in personal network sizes are a contributing factor to the differences in centralities. However, the strong level of homophily between owls and the asymmetric absence of homophily between larks is not directly explained by degrees or centralities. A possible candidate explanation is the need to synchronize social activities, together with common social norms. Social gatherings, including student parties, often take place in the late hours of the day. Because social ties are created and maintained in such events that are more suitable for owls, it is perhaps not surprising that there is a bias in favor of the evening-active chronotype. In contrast, early-morning social gatherings are not common, and it is therefore possible that those with a very early chronotype tend to spend more of their time alone, and get to interact with fewer people. 

Additionally, there is a possible connection between chronotype-related homophily to social influence. It has been shown in previous work that the evening-active chronotype is more susceptible to disease and health problems than the morning-active one~\cite{adan2012circadian, merikanto2014evening, yu2015evening}. Further, it has been suggested in several papers that social networks have an influence on health through social ties, and that healthy or unhealthy behaviors spread on social networks~\cite{christakis2007spread, christakis2008collective, aral2009distinguishing, aral2012identifying, aral2017exercise}. At the same time, our result point out that there is an increased number of social ties between individuals of the evening-active chronotype which is more prone to disease; this would make it possible that increased exposure to unhealthy behavioral patterns of others of the same chronotype is a contributing factor. Whether these issues are related poses an interesting problem for future research. 
%[One paragraph on homophily, Christakis stuff, and posing an open question]

 %here are suggestions (Christakis papers. e.g. “The Spread of Sleep Loss Influences Drug Use in Adolescent Social Networks”) that our social network influences our health and unhealthy behaviors of each node can influence neighboring nodes.
%We show homophily in chronotypes, and it is shown that certain chronotype (owls) are more prone to many diseases. Our findings can be a key to the reason why there we see that for example neighboring nodes of overweight people are also overweight (because there is more chance that the neighbors are also same chronotype).

%; Talayeh's points regarding social influence and spreading as an open question

Regarding determining chronotypes, the most common way has been to use questionnaires.%this sentence is not quite right, the physiological measurements measure endogenous circadian rhythms of physiological variables, whereas questionnaires are designed for epidemiological studies, and are meant to measure chronotypes 
It has been suggested that questions related to sleeping and waking up habits can explain most of the variance in distinguishing chronotypes~\cite{adan2012circadian}. This is the rationale behind our method: screen-on events are used as proxy of activity, and early-morning and night-time activity levels determine chronotype. The specific time intervals have been determined from data as time ranges where population-level average activity is only started to increase from very low values (mornings), or is about to cease for the night. We have also taken into account differences between weekend and weekday patterns, focusing on weekdays where the students have approximately similar schedules; the literature suggests that especially owls develop a sleep deficit during the week and their weekend patterns may differ as they compensate for this deficit. 

The data we use were not originally collected for determining or studying the chronotypes of individuals. As we used the data for this purpose retrospectively, there is no way of validating the obtained chronotypes against common methods such as questionnaires; this would require an experiment of its own. However, it is reasonable to assume that if an individual in our data persistently displays above-average levels of early-morning activity and below-average levels of night-time activity, this pattern would be captured by questionnaires developed for assessing chronotypes, with a similar outcome. In general, the possibility of retroactively determining people's chronotypes from time-stamped data recorded for other purposes opens a lot of possibilities for research, especially together with our observations that chronotypes are not independent of social network structure (or vice versa): there are many datasets available to researchers from time-stamped mobile telephone calls~\cite{blondel2015survey,saramaki2015seconds} to email records that allow for retrospectively determining chronotype and reconstructing social networks~\cite{aledavood2015cycles}. 

\section{\label{section_methods}Materials ans Methods}
\subsection{Experimental data and its filtering}
In this work, we used data from a large-scale data collection 
study, in which $1000$ identical mobile phones were distributed among students at Technical University of Denmark (DTU) in $2013$. This data collection experiment was designed to measure various aspects of social behavior and human dynamics. In the study, smartphone apps were used to collect data with high temporal resolution from various mobile phone sensors. These data were augmented with additional questionnaires that all students in the study filled out. The data collection continued for more than a year; all participants were, however, not active throughout the whole period. A detailed description of this dataset, types of collected data, and research envisioned by means of the dataset are described in~\cite{stopczynski2014DTU}. In the present paper, data from weeks $2-51$ of year $2014$ are used. Weeks $1$ and $52$ are excluded, because the former starts in the end of $2013$ and the latter coincides with end-of-the-year holidays, which may result in untypical activity patterns. The number of participants that used their phone during this year is $N=804$. We apply filters to the data to only include participants who use their phone actively and exclude those who have very little activity or are inactive for a part of the year. The inclusion criteria were: 1) the participant should be active on $80\%$ of the days, 2) on every week included in the study (weeks $2-51$), the participant should at least have $280$ screen-on and screen-off events. This means that the phone screen should turn on at least $20$ times per day on average. After applying these filters, the total number of participants included for further analysis is $N=222$. The data and the associated code used for this paper will be made available to researchers upon request.

\subsection{Computing the activity patterns of students}
For computing the activity patterns of individuals, one week is divided to one-hour bins. As a result, we have a total of $7 \times 24 = 168$ bins. For each of the $222$ active participants, we go through all screen-on events in the selected study period and assign each event to one of the bins. This way, all events for one person throughout the year are aggregated. These aggregated counts are then normalized so that their values sum to unity. This results in a weekly activity pattern for each person.

\subsection{Identifying chronotypes from activity patterns}
Weekly activity patterns of individuals are used as the basis of assessing chronotypes. Only weekdays from Monday to Thursday were used in the analysis. The reason for this is that the literature suggests that individuals may have different behaviors during working days and weekends, and the extent of these differences may vary from one chronotype to another~\cite{wittmann2006social, roenneberg2012internal}. 
%"The largest differences between M-types and E-types in their answers on a morningnesseveningness questionnaire are to be found on items which relate to their activities in early and late hours." from: ~\cite{kerkhof1985InterIndividual}

We find the population's average weekly pattern by computing the average of the patterns of all $222$ participants. Then, for each person, their weekly pattern is compared with the population average in the early hours of the day ($5$ AM-$7$ AM) as well as the late hours (midnight-$2$ AM). To label an individual as a lark, her pattern should lie above the population average in the early hours, and in addition, in the late hours the difference between the individual's pattern and the population average should be less than $0.0033$. The same applies to identifying owls: at late hours of the night, their pattern should be above the population average and in the early morning hours the difference of the individual's rhythm and the population average should be less than $0.002$. The numerical thresholds for the differences between activity patterns are selected so that 20\% of the population is labeled as larks and 20\% as owls. These percentages have been chosen to match the literature~\cite{roenneberg2007epidemiology, levandovski2013ChronotypeReview}. The time intervals for early-morning and night-time hours have been selected based on the times when the population average rhythm falls at night and when it begins to rise again in the morning.

%Previous work have shown that one can learn from temporal features and rhythms of a system by looking at system averages of activity level\cite{aledavood2015cycles, yasseri2012circadian}. 

%The chronotype extraction task could be tackled with several different tools, including principal and independent component analyses, factor analysis, topic modeling as well as some functional data analysis methods. However, we can get similar results using a simple and straightforward that we implemented. The results from this method is compared to the best results from non-negative matrix factorization in the method in the SI section, and little difference can be seen across methods. 
%In addition to NMF, we did also experiment with other methods (mentioned above), but found the results obtained with NMF most interesting and informative.
%In essence, we chose NMF for its interpretability (e.g.\ compared to component analysis methods which do not have the non-negativity constraint) and its conceptual simplicity (over more involved methods such as topic modeling and functional data analysis methods).
%Also, the phone screen activity data that we analyze is non negative, which further makes NMF a natural choice.

\subsection{Reconstructing the social network of students}

Students in the study were all from the same university and mostly began their studies at the same time. To construct the students' social network, all individuals in the study who had communication events during the year 2014 were selected ($N=776$). In panel E of Fig.~\ref{fig:network}, the social links between students in the study is depicted. Out of the $776$ students in the network, $218$ had an identified chronotype; $4$ students with an identified chronotype were not part of the network because they did not have any calls or text messages. For all panels in Fig.~\ref{fig:network}, incoming and outgoing calls and text messages were used. However, only those social ties were included that were associated with at least one event in both directions (incoming and outgoing), to assure that a tie between two individuals is representative of a social relationship.

The dataset also contains calls and text messages between study participants and people from outside the study cohort. This makes it possible to build a more comprehensive picture of each individual's personal network based on communication events. For the $218$ students with known chronotypes, personal network are built from all outgoing calls and texts. There are in a total of $13608$ social links for the $218$ students; again we only kept social links which were active at least once in each direction. The properties of the personal network shown in Fig.~\ref{fig:chronotype_behaviours} are based on outgoing communication only. 

\subsection{Centrality measures and $k$-shells}

The notion of centrality of a network node can be defined in several ways. In this work, different centrality measures are used: betweenness centrality, eigenvector centrality, and closeness centrality. Betweenness centrality is a proxy of the importance of flows in the network, and it is defined on the basis of the number of shortest paths that a node is a part of; the more shortest paths go through a node, the higher its betweenness centrality. The betweenness centrality $C_b(i)$ for node $i$ can be formulated as 
\[
C_b(i) = \sum_{j<k}\frac{\sigma_{jk}(i)}{\sigma_{jk}}, 
\]
where $\sigma_{jk}$ accounts for the total number of shortest paths from node $j$ to node $k$ and ${\sigma_{jk}(i)}$ is the number of those paths which go through node $i$.

Closeness centrality quantifies how far each node is from all other nodes in the network. The closeness centrality $C_c(i)$ for node $i$ is defined as 
\[
C_c(i) = \frac {1}{\langle l_i \rangle} = \frac {N-1}{\sum_{i\neq j} d_{ij}},
\]
where $\langle l_i \rangle$ is the average length of shortest paths of node $i$ to all other nodes in the network, $d_{ij}$ is the shortest path between nodes $i$ and $j$, and $N$ is the total number of nodes in the network.

Eigenvector centrality is an iterative centrality measure that does not only depend on how well-connected each node is, but also takes into account the centrality of its neighbors, such that a node with a small number of central neighbors may outrank one with more less central neighbors~\cite{newman2008NetworkMathematics}. Therefore, it measures how well-connected each node is to other well-connected nodes. Formally, eigenvector centrality of a node is defined as the corresponding element of the eigenvector of the network's adjacency matrix that corresponds to its largest eigenvalue.

The other measure we use is the core number which is defined based on the concept of $k$-cores. $k$-cores are maximal subgraphs in the network where all nodes are connected to other nodes in the subgraph with at least $k$ links. For a node, the core number is the largest value of $k$ for $k$-cores that contains the node~\cite{carmi2007kshell}.

\section*{Acknowledgment}
The authors thank the following people for their help and support at different stages of this work: Ilkka Kivim\"aki, Bjorn Sand Jensen, Rasmus Troelsg\r{a}rd, Richard Darst, Radu Gatej, and Onerva Korhonen. TA acknowledges the support of EIT Digital doctoral school for her visit to DTU. JS and TA acknowledge support from the Academy of Finland, project DigiDay, No. 297195.

\bibliography{refs}
\end{document}